# Calculated Electronic and Related Properties of Wurtzite and Zinc Blende Gallium Nitride (GaN)

Yacouba Issa Diakité[1], Sibiry D. Traoré[1], Yuriy Malozovsky[2], Bethuel Khamala[2], Lashounda Franklin[2], Diola Bagayoko[2]

[1]Departement of Studies and Research (DSR) in Physics, College of Science and Technology (CST), University of Science, Techniques, and Technologies of Bamako (USTTB), Bamako, Mali

[2]Southern University and A&M College in Baton Rouge (SUBR), Louisiana, USA

**All Correspondence:**

**Diola Bagayoko**

**Phone: +1225 205 7482**

**Email: bagayoko@aol.com**

**Postal Address: Department of Physics, William James Hall, Southern University and A&M College, P.O. Box 11776, Baton Rouge, Louisiana 70813, USA.**

## Abstract

We report calculated, electronic and related properties of wurtzite and zinc blende gallium nitrides (w-GaN, zb-GaN). We employed a local density approximation (LDA) potential and the linear combination of atomic orbital (LCAO) formalism. The implementation of this formalism followed the Bagayoko, Zhao, and Williams (BZW) method, as enhanced by Ekuma and Franklin (BZW-EF). The calculated electronic and related properties, for both structures of GaN, are in good agreement with corresponding, experimental data, unlike results from most previous ab initio calculations utilizing a density functional theory (DFT) potential. These results include the electronic energy bands, the total and partial densities of states (DOS and pDOS), and effective masses for both structures. The calculated band gap of 3.29 eV, for w-GaN, is in agreement with experiment and is an average of 1.0 eV larger than most previous ab-initio DFT results. Similarly, the calculated band gap of zb-GaN of 2.9 eV, for a room temperature lattice constant, is the ab-initio DFT result closest to the experimental value.

## I. Introduction

Gallium nitride is a wide-band-gap semiconductor with applications in optical devices (light-emitting diodes, lasers, detectors) in the near UV region of the electromagnetic spectrum, electronic devices for high-power, and in high-frequency and high-temperature applications [1]. GaN exists in two allotropic forms. The wurtzite structure is the thermodynamically stable one at ambient temperature and pressure; different experiments found the room temperature band gap of w-GaN to be a direct one, with measured values of 3.26[2], 3.39 [3], and 3.41 [4] eV. The latter two values were obtained from films known to have high electron concentrations, i.e., $1-5 \times 10^{19}/cm^3$ or higher for the last one. We could not determine the source of the difference between the first and the last two values. Quantum confinement effects can explain slightly higher gaps for thinner films. Also, a Burstein-Moss effect can, with very high carrier concentrations, lead to larger band gaps. Powell and co-workers [4] discussed experimental results for the band gap of zinc blende GaN. While these authors measured a room temperature direct gap of 3.21 0.02 eV, they noted other experimental ones of 3.206 eV and 3.26 eV. Unlike the case of the above experimental results that basically agree with each other, for each of the phases of GaN, theoretical ones seem to disagree among themselves and with experiment as shown below.

Even though many theoretical calculations of the electronic properties of wurtzite gallium nitride (w-GaN) have been reported in the literature, obvious discrepancies exist not only between results from different calculations, but also between theoretical and corresponding experimental values of several electronic properties, with emphasis on the band gap. Table 1a contains a listing of several calculations and of the resulting band gaps for w-GaN. Except for the first one, which was obtained with an early version of the method used here, nine (9) previous ab-initio LDA calculations reported band gaps between 1.9188 eV and 2.5181 eV [5]. The latter, larger value is 0.74 eV smaller that the smallest, reported experimental one. Similarly, some eight (8) ab-initio calculations using a generalized gradient approximation (GGA) potential found even smaller band gaps ranging from 1.68 eV to 1.95 eV [6, 7, 8]. A GGA plus U parameter calculation led to a gap of approximately 2.5 eV [9]. Some Green function (G) and dressed Coulomb (W) approximation (GW) calculations, for quasi particle energies, yielded theoretical gaps of 3.448 and 3.5 eV [9, 10], in basic agreement with the experimental value of 3.4 eV. GW calculations are out of the scope of DFT which is ground state theory. Clearly, the above theoretical results directly point to the need for further work, particularly in the framework of ab-initio DFT calculations.

**Table 1a.** Previous, calculated, fundamental band gaps ($E_g$, in eV) of wurtzite gallium nitride (w-GaN), along with pertinent lattice constants in Angstroms, the u parameter, and measured band gaps.

| Computational Formalism and Method | Potentials (DFT and Others) | a(Å) | c(Å) | u | Eg (eV) |
|---|---|---|---|---|---|
| LCAO BZW | LDA | 3.16 | 5.125 | 0.377 | 3.2[d] |
| PP plane wave | LDA | 3.162 | 5.142 | 0.377 | 2.04[e] |
| Ab initio speudopotential | LDA | 3.126 | 5.119 | 0.377 | 2.3[f] |
| FP-LAPW | LDA | 3.1555 | 5.1529 | | 2.09[h] |
| FP-LAPAW and pseudopotential | LDA | 3.163 | 5.140 | | 2.08[i] |
| FP-L/APW+LO | LDA | 3.157 | 5.14 | | 2.10[k] |
| FLAPW | PB96 LDA | 3.2301 | 5.0311 | 0.3769 | 1.9188[c] |

| FLAPW | EV93 LDA | 3.2301 | 5.0311 | 0.3769 | 2.5181[c] |
| PAW | PP-LDA | 3.17 | 5.15 | 0.377 | 2.1[l] |
| APW+LO | GGA | 3.1904 | 5.1908 | | 1.83[m] |
| FP-L/APW+LO | GGA | 3.224 | 5.25 | | 1.691[k] |
| Ab-initio pseudopotential | GGA | 3.26 | 5.18 | | 1.95[n] |
| FP-LAPW | GGA | 3.2209 | 5.2368 | 0.3780 | 1.768[o] |
| FP-LAPAW and pseudopotential | GGA | 3.216 | 5.243 | | 1.68[i] |
| FP-LAPW | GGA | 3.2241 | 5.2488 | | 1.68[h] |
| PAW | GGA | 3.213 | 5.240 | | 1.774[j] |
| FLAPW | GGA | 3.159 | 5.147 | 0.3765 | 1.92[g] |
| PAW | GGA+U | 3.154 | 5.135 | | 2.489[j] |
| Ab initio speudopotential | GW | 3.126 | 5.119 | 0.377 | 3.5[f] |
| PW | GW | | | | 3.448[j] |
| Experiment | | | | | |
| Experimental values | Optical-absorption measurements at room temperature | | | | 3.39[a] |
| | Optical-absorption measurements at room temperature | | | | 3.41± 0.02[b] |
| | SXE and SXA measurements | | | | 3.4 ± 0.2[m] |

[a]Reference[3], [b]Reference[4], [c]Reference[5], [d]Reference[16], [e]Reference[25], [f]Reference[10], [g]Reference[26], [h]Reference[6], [i]Reference[7], [j]Reference[9], [k]Reference[27], [l]Reference[28], [m]Reference[29], [n]Reference[8], [o]Reference[30].

In the first column of this table and the next one, the following meanings hold for the abbreviations: pseudopotential (PP); plane wave (PW), augmented plane wave (APW), linearized APW (LAPW), full potential LAPW (FL-LAPW), local orbital (lo)

The above unsatisfactory picture for the description of w-GaN by ab initio DFT calculations is similar to that for the zinc blende structure. Indeed, the content of Table 1b shows that the calculated band gaps for zinc blende GaN (zb-GaN) are between 1.8 eV and 2.1eV [11, 10] for ab-initio, local density approximation calculations. The latter, higher value is about 1 eV smaller than the experimental gap. Five (5) different first principle GGA calculations produced even smaller gaps between 1.51 eV and 1.92 eV [12, 13] for zb-GaN. An empirical pseudo potential method calculation, which entails extensive fitting, reported a gap of 3.38 eV [14] while a generalized DFT calculation, with corrections to the excited energies, obtained 3.910 eV [15]. For a lattice constant usually small (4.35 Å as compared to the 4.45 – 4.55 Å), a GW calculation obtained a gap of 3.1 eV [10] that is in agreement with experiment.

**Table 1b.** Previous, calculated, fundamental band gaps ($E_g$, in eV) of zinc blend gallium nitride (zb-GaN), along with the pertinent lattice constants, in Angstroms, and measured room temperature band gaps.

| Computational Formalism and Method | Potentials (DFT and Others) | a(Å) | $E_g$ (eV) |
|---|---|---|---|
| | | | |
| FP-LAPW | LDA | 4.52 | 1.80[p] |
| PP-PW | LDA | 4.460 | 1.89[e] |
| FP-LAPW | LDA | 4.461 | 1.9[q] |
| Ab initio speudopotential | LDA | 4.42 | 2.1[f] |
| PP-PW | LDA | 4.47 | 2.04[t] |
| PP-PW | LDA | 4.47 | 1.98[v] |
| FP-L/APW+LO | LDA | 4.461 | 1.93[k] |
| FP LAPW | LDA | 4.475 | 1.8[w] |
| | | | |
| LDA-1/2 | LDA + correction | 4.458 | 3.3[x] |
| EPM | LDA with fitting | 4.50 | 3.38[u] |
| GDFT | Quasi particle | 4.52 | 3.910[y] |
| | | | |
| FP-L/APW | GGA | 4.55 | 1.521[k] |
| Pseudopotential method | GGA | 3.20 | 1.8[n] |
| PP-PW | GGA | 4.50 | 1.92[v] |
| FP-LAPW | GGA | 4.55 | 1.51[q] |
| FLAPW | GGA | 4.468 | 1.83[g] |
| FLAPW | LDF-GGA | 4.552 | 1.72[s] |
| Ab initio speudopotential | GW | 4.35 | 3.1[f] |
| Experiment | | | |
| Experimental | Modulated photoreflectance measurements at room temperature | | 3.231± 0.008[r] |
| | Optical-absorption measurements at room temperature | | 3.21± 0.02[b] |

[b]Reference[4], [e]Reference[25], [f]Reference[10], [g]Reference[26], [k]Reference[27], [n]Reference[8], [p]Reference[11], [q]Reference[12], [r]Reference[31], [s]Reference[32], [t]Reference[33], [u]Reference[14], [v]Reference[13], [w]Reference[34], [x]Reference[35], [y]Reference[15]

The above overview of previous theoretical results, with the general disagreement between calculations, on the one hand, and between theoretical values and corresponding, experimental ones, on the other hand, shows the need for further work. The aim of the present effort is not only to obtain calculated, electronic energies in basic agreement with experiment, but also to provide an alternative explanation, other than the presently accepted ones, for some of the limitations of previous, ab initio density functional (DFT) calculations. *It is critical to note that we distinguish "limitations of DFT calculations" from "intrinsic limitations or shortfalls of DFT."* An early version of our distinctive method led to a direct band gap of 3.2 eV[16] for w-GaN. Known as the Bagayoko, Zhao, and Williams (BZW) method, it has recently been enhanced by the work of Ekuma and Franklin to become the BZW-EF method. The closeness of the above BZW result of 3.2 eV to the experimental ones (3.26 – 3.41 eV), in contrast to other ab-initio DFT ones in table I, suggests that the BZW-EF method could lead to improvements in the theoretical description of w-GaN and of zb-GaN. The rest of this paper is organized as follows. After this introduction in Section I, our approach, including the BZW-EF method and related computational details, are given in Section II. The results of our self-consistent calculations are presented and discussed in section III for both w-GaN and zb-GaN. We conclude with a summary in Section IV.

## II. Method

Our calculations employed the Ceperley and Alder [17] local density approximation (LDA) potential, as parameterized by Vosko, Wilk, and Nusair [18], and the linear combination of atomic orbitals. The radial parts of these orbitals included Gaussian functions. We utilized a program package developed and refined over decades [17, 19] at the US Department of Energy's Ames Laboratory, Ames, Iowa. Our calculations are non-relativistic and are performed using an experimental room temperature lattice constant for both structures of GaN. The distinctive feature of our approach resides in our implementation of the (BZW-EF) method consisting of concomitantly solving self-consistently two coupled equations. [20] One of these equations is the Schrödinger type equation of Kohn and Sham [21], referred to as the Kohn-Sham (KS) equation. The second equation, which can be thought of as a constraint on the KS equation, is the one giving the ground state charge density in terms of the wave functions of the occupied states.

Our solid state calculations begin with those for the atomic or ionic species which are present in the material. In the case of GaN, taking into account previously found [16] charge transfer, we performed ab initio, self consistent calculations of electronic properties of $Ga^{+1}$ and $N^{-1}$. These calculations produced the energy levels of these ions and the corresponding wave functions. While there is some flexibility in the choice of the exponents, these calculations have to conserve electronic charges as much as possible. Specifically, given the quintessential role of the charge density in DFT, the charges on these species have to be as close to the applicable integers as possible, i.e., 30 and 8 for $Ga^{+1}$ and $N^{-1}$. We did not accept any discrepancy smaller than $10^{-3}$. The wave functions obtained in the above calculations are used in the linear combination of atomic orbitals to construct the wave functions for the GaN.

According to the BZW-EF method, calculations for solids start with a small basis set that is no smaller than the minimum basis set. The latter is simply the basis set just large enough to account for all the electrons in the system. The first self consistent calculation for the solid is performed using this small basis set. The method requires that other calculations follow, with successively larger basis set as explained above and below. For the second calculation, the initial basis set is augmented by adding one orbital for an excited state. Depending on the s, p , d, or f character of this orbital, this addition means that the total number of basis functions, including angular symmetry and spin, increases by 2, 6, 10, or 14, respectively. In the initial BZW method, the addition of orbitals followed the increasing level of the excited energies of the atomic or ionic species in the solid. In the BZW-EF method, followed in this work, this rule is relaxed; for a given principal quantum number, p, d, and f orbitals (if applicable) are added before the corresponding s orbital. The work of Ekuma et al. [22] and Franklin et al. [23] provided the reasons for this approach: they consist of offering optimal flexibility to the electronic cloud, *of the valence states*, to reorganize itself in the solid. For valence electrons, polarization has primacy over spherical symmetry in solids [23].

Following the above method of augmenting the basis set from one calculation to the next, we performed several self consistent calculations. Except for the first one, the occupied energies of each calculation are compared graphically and numerically to those of the calculation immediately preceding it. In general, some occupied energies of a calculation are lower than corresponding ones from the calculation immediately preceding it. This process continues until a calculation, i.e., N, is found to have the same occupied energies as the one preceding it, i.e., (N-1). This superposition of the occupied energies signifies that the energies from Calculation (N-1) are minimal ones, as required by the very derivation of DFT. Calculation (N+1) is performed and its occupied eigenvalues are compared to those of Calculation N. If, instead of perfect superposition, some occupied energies of Calculation (N+1) are found to be lower than those of Calculation N, then the above minimal energies were relative (or local) ones as opposed to being the absolute minima. *The process continues until the absolute, minimal values of the occupied energies are found*. The BZW-EF method concludes the reaching of the absolute minima of the occupied energies when three (3) successive and consecutive calculations produce identical values of these occupied energies, within the computational uncertainties of 5 meV. The first of these three calculations, i.e., the one with the smallest basis set, provides the DFT description of the system and the corresponding basis set is called the *optimal basis* set. For both structures of GaN, calculation III is the one giving the DFT description of the material.

When Calculation (N-1) leads to the absolute minimal values of the occupied energies, Calculations N, (N+1), (N+2), etc., do not modify the occupied energies. In general, with the BZW-EF method, Calculation N and (N-1) also produce the same values for (up to +10 eV) for the lowest, unoccupied energies. However, Calculations (N+1) and those with much larger basis sets that contain the optimal one lead to some unoccupied energies that are lower than the corresponding ones obtained with the optimal basis set. While "the larger the basis set, the better" is a common rule, it leads to errors due to the Rayleigh theorem, as explained by Bagayoko et al. [24], Ekuma et al. [22], and particularly by Franklin et al. [23]. The latter authors [Franklin et al.] provided pertinent illustrations of the effects of much larger basis sets and of imposing "5s" symmetry on the valence electrons for ZnO. This extra-lowering of some unoccupied energies, after the attainment of the absolute minima of the occupied ones, is a mathematical artifact stemming from the Rayleigh theorem. This well defined "basis set and variational effect" is different from the vague "basis set effect." It can be invoked only after verifiably attaining the absolute minima of the occupied energies.

For the sake of completeness, and particularly for further reference in the discussion section, we state here the Rayleigh theorem [36]: Let an eigenvalue equation, $H\Psi = \lambda\Psi$, be solved by the LCAO method with N functions and (N+1) functions, such that the N functions are entirely included in the (N+1) functions; and let the resulting eigenvalues be ordered from the smallest to the largest; then, $\lambda_i^{(N+1)} \leq \lambda_i^N \; for \; i \leq N$. In other words, this theorem says the following : When the same eigenvalue equation is solved by the procedure of the linear combination of functions M and (M +1) functions , with the M functions entirely included in the (M+1) functions, then the ordered eigenvalues obtained with (M +1) functions are smaller than or equal to the corresponding ones obtained with M functions.

Computational details germane to the replication of our work follow. Wurtzite GaN belongs to the $C_{6v}^4$ group. We used the experimental room temperature lattice constants of a =3.16Å and c=5.125Å with a "u" parameter of 0.377,[10] where u is the distance between the Ga plane and its nearest-neighbor N plane in the unit of cell. The "atomic" wave functions of the ionic states of $Ga^{+1}$ and $N^{-1}$ were obtained, as noted earlier, from self consistent calculations. They provided the orbitals in the LCAO calculations for the solids. The radial parts of these functions were expanded in terms of Gaussian functions. A set of even tempered Gaussian exponents was employed with a minimum of 0.3 and a maximum of $0.5585 \times 10^5$ in atomic units $Ga^{1+}$. We used 19 Gaussian functions for the s and p states and 16 for the d states of $Ga^{1+}$. The two largest exponents are ignored for the d state. Similarly, a set of even-tempered Gaussian exponents was utilized to describe $N^{-1}$, with a minimum of 0.22 and a maximum of $0.5254 \times 10^5$. Both the s and p functions were expanded in terms of 13 Gaussian orbitals. The specific Gaussian exponents employed in our calculations are listed below in Tables 2a and 2b for $Ga^{1+}$ and $N^{1-}$, respectively. A mesh of 24k points, with proper weights in the irreducible Brillouin zone of the wurtzite structure, was used in the iterations for self consistency. The computational error for the valence charge was about 0.0028 for 52 electrons, or $5.4 \times 10^{-5}$ per electron. The self-consistent potentials converged to a difference around $10^{-5}$ for two consecutive iterations, after about 60 iterations. The total number of iterations varied slightly with the input potentials.

Zinc-blende GaN is a member of the III-V family. The "atomic" wave functions for $Ga^{1+}$ and $N^{1-}$ were obtained as described above for w-GaN. We employed a room temperature lattice constant of 4.50 Å. A set of even-tempered Gaussian exponents was employed for $Ga^{1+}$ and $N^{1-}$ are as described above for the wurtzite structure. A mesh of 60k points in the irreducible Brillouin zone, with proper weights, was used in the iterative process. The computational error for the valence charge was about 0.000205 for 26 electrons, or $7.9 \times 10^{-6}$ per electron. The self-consistent potentials converged to a difference around $10^{-5}$ between two consecutive iterations, after about 60 iterations.

**Table 2a.** Gaussian Exponents for $Ga^{1+}$

| | | | |
|---|---|---|---|
| 0.29999993E+00 | 0.50844893E+00 | 0.86173456E+00 | 0.14604937E+01 |
| 0.24752887E+01 | 0.41951938E+01 | 0.71101409E+01 | 0.12050481E+02 |
| 0.20423518E+02 | 0.34614394E+02 | 0.58665518E+02 | 0.99428086E+02 |
| 0.16851372E+03 | 0.28560212E+03 | 0.48404708E+03 | 0.82037757E+03 |
| 0.13904006E+04 | 0.23564929E+04 | 0.39938551E+04 | 0.67689059E+04 |
| 0.11472146E+05 | 0.19443338E+05 | 0.32953155E+05 | 0.55850000E+05 |

**Table 2b.** Gaussian Exponent for $N^{-1}$

| | | | |
|---|---|---|---|
| 0.21999986E+00 | 0.50229697E+00 | 0.11468291E+01 | 0.26184053E+01 |
| 0.59782631E+01 | 0.13649388E+02 | 0.31163867E+02 | 0.71152390E+02 |
| 0.71152390E+02 | 0.37090766E+03 | 0.84684507E+03 | 0.19334908E+04 |
| 0.44144873E+04 | 0.10079023E+05 | 0.23012118E+05 | 0.52540565E+05 |

Using the LCAO formalism and the LDA potential, along with the computational details provided above, we followed the BZW-EF method to perform ab initio, self consistent calculations of electronic energies and related wave functions for w-GaN and zb-GaN. With these energies and wave functions for the calculations with the optimal basis sets, we have additionally calculated total (DOS) and partial (pDOS) densities of states for both structures. These results are presented and discussed in the next section.

## III. Results and Discussions

### a. Results

We first present the successive calculations that led to the absolute minima of the occupied energies for w-GaN and zb-GaN. We then present the electronic energy bands resulting from the calculations with the optimal basis sets, followed by the total and partial densities of states and calculated effective masses. Using the figures for the total and partial densities of states, some features of the electronic properties are pointed out. The last results to be presented are the bulk properties of zb-GaN.

#### a.1. The Successive Calculations of the BZW-EF Method

Tables 3a and 3b below show the successive calculations, required by the BZW-EF method, that led to the results presented here. For both w-GaN and zb-GaN, Calculation III is the one that led to the absolute minima of the occupied energies. These calculations provide the DFT description of the systems under study. Our calculated, direct, band gap, at the $\Gamma$ point, is 3.289 ( 3.29) eV for w-GaN; it is in agreement with experimental values of 3.26 eV and only 0.11 eV smaller than 3.41 eV that was obtained with large carriers concentration. This LDA BZW-EF result is a drastic improvement over most previous ab initio DFT calculations, with LDA or GGA potential, reviewed in the introduction. The only exception is the work of Zhao et a. that employed the BZW method. For zb-GaN, the calculated, direct band gap of 2.896 eV (2.90 eV) is practically in agreement with experiment, particularly as compared to previous ab initio DFT results. The electronic energy bands from these calculations with the optimal basis sets are presented in the next sub-sections.

**Table 3a.** Successive calculations of the BZW-EF method for w-GaN (Calculations I –V). The lattice constants are a = 3.16Å and c = 5.125Å, with u = 0.377. Calculation III led to absolute minima of the occupied energies; the corresponding basis set is the optimal basis set. The calculated, direct band gap, at $\Gamma$, is 3.289 eV.

| Calculation number | Orbitals for $Ga^{1+}$ | Orbitals for $N^{1-}$ | Number of functions | Band Gap (eV) |
|---|---|---|---|---|
| Calc I | $3s^2 3p^6 3d^{10} 4s^2 4p^0$ | $2s^2 2p^4$ | 68 | 3.213 |
| Calc II | $3s^2 3p^6 3d^{10} 4s^2 4p^0\ 4d^0$ | $2s^2 2p^4$ | 88 | 3.316 |
| **Calc III** | $\mathbf{3s^2 3p^6 3d^{10} 4s^2 4p^0 4d^0\ 5p^0}$ | $\mathbf{2s^2 2p^4}$ | **100** | **3.289** |
| Calc IV | $3s^2 3p^6 3d^{10} 4s^2 4p^0 4d^0 5p^0 5d^0$ | $2s^2 2p^4$ | 120 | 3.322 |
| Calc V | $3s^2 3p^6 3d^{10} 4s^2 4p^0 4d^0 5p^0 5d^0 5s^0$ | $2s^2 2p^4$ | 124 | 2.933 |

**Table 3b.** Successive calculations of the BZW-EF method for zb-GaN (Calculations I –V). The utilized lattice constant is a = 4.50 Å, at room temperature. Calculation III led to absolute minima of the occupied energies; the corresponding basis set is the optimal basis set. The calculated, direct band gap, at Γ, is 2.896 eV

| Calculation number | Orbitals for $Ga^{1+}$ | Orbitals for $N^{1-}$ | Number of functions | Band Gap in eV |
|---|---|---|---|---|
| Calc I | $3s^23p^63d^{10}4s^24p^0$ | $2s^22p^4$ | 34 | 2.798 |
| Calc II | $3s^23p^63d^{10}4s^24p^04d^0$ | $2s^22p^4$ | 44 | 2.887 |
| **Calc III** | $\mathbf{3s^23p^63d^{10}4s^24p^04d^0\,5p^0}$ | $\mathbf{2s^22p^4}$ | **50** | **2.896** |
| Calc IV | $3s^23p^63d^{10}4s^24p^04d^05p^05d^0$ | $2s^22p^4$ | 60 | 2.964 |
| Calc V | $3s^23p^63d^{10}4s^24p^04d^05p^0\,5d^05s^0$ | $2s^22p^4$ | 62 | 2.573 |

### a.2 Electronic Energy Bands of w-GaN and zb-GaN

Figures 1a and 1b show the calculated band structures of w-GaN and zb-GaN, respectively, as obtained from calculations III (full line) and IV (dashed line). The perfect superposition of the *occupied energies continues for unoccupied ones up to 10 eV*. Some basic features of these bands are noted below, in connection with the DOS and pDOS data and of the listing of calculated energies at some high symmetry points in the Brillouin zone.

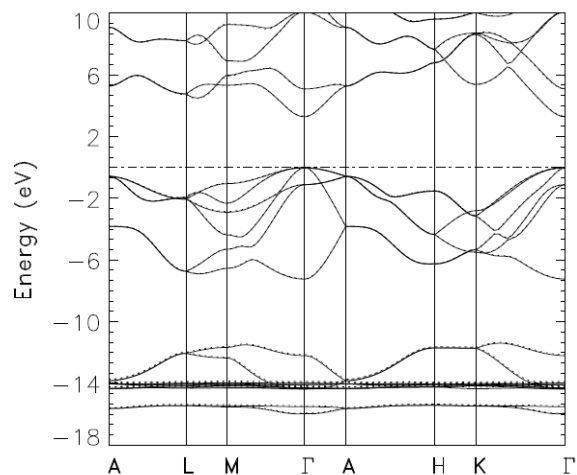

**Figure1a.** Calculated band structures of wurtzite gallium nitride (w-GaN), as obtained by the BZW-EF method from calculations III(_, full line)) and IV(_ _, dashed line). The calculated, direct band gap is 3.289 (3.29) eV. *For the shown bands (occupied and unoccupied), the results from Calculations III and IV are the same, within the computational uncertainties estimated at 0.005eV, as attested to by the perfect superposition of the bands.* The Fermi energy ($E_F$) has been set to zero.

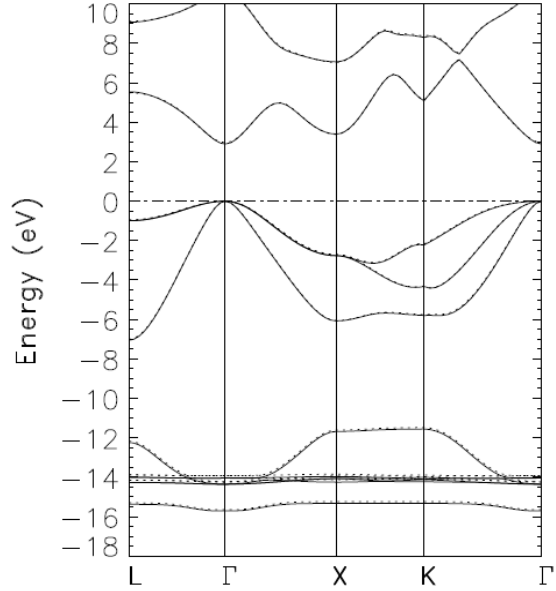

**Figure1b.** Calculated band structures of zinc blende gallium nitride (zb-GaN), as obtained by the BZW-EF method from calculations III (__) and IV(---). The calculated, direct band gap is 2.896 (2.9) eV. *For the bands shown (occupied and unoccupied), the results from Calculations III and IV are the same, within the computational uncertainties estimated at 0.005eV, as attested to by the perfect superposition of the bands.* The Fermi energy ($E_F$) has been set to zero.

### a.3 Densities of States and Eigenvalues at High Symmetry Points

Figures 2a and 2b respectively show the total densities of states (DOS) for w-GaN and zb-GaN, as obtained from the bands respectively shown in Figures 1a and 1b. Figures 3a and 3b exhibit the partial densities of states (pDOS) for w-GaN and zb-GaN, as obtained from the bands in Figures 1a and 1b, respectively. Farther below, we discuss the pDOS for w-GaN in comparison to some available, experimental data on hybridization patterns.

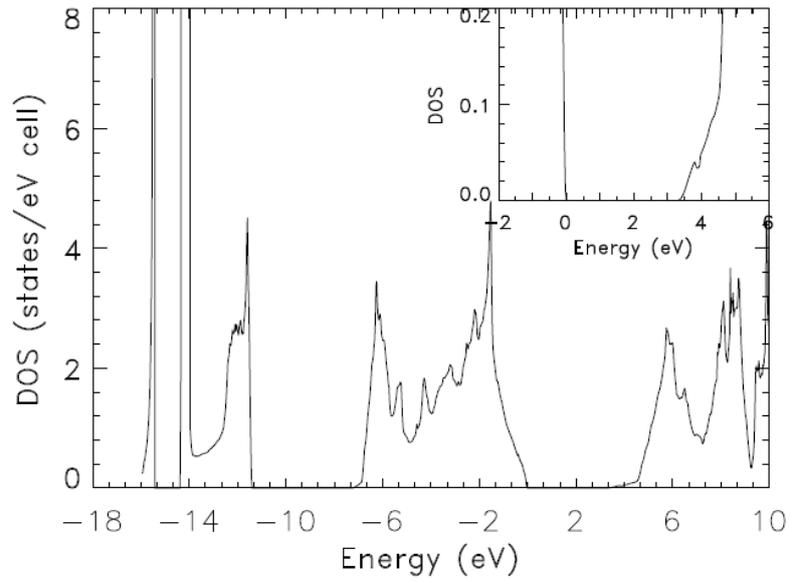

**Figure 2a.** Calculated, total density of state (DOS) for w-GaN, as derived from the bands from calculation III, as shown in Fig.1a

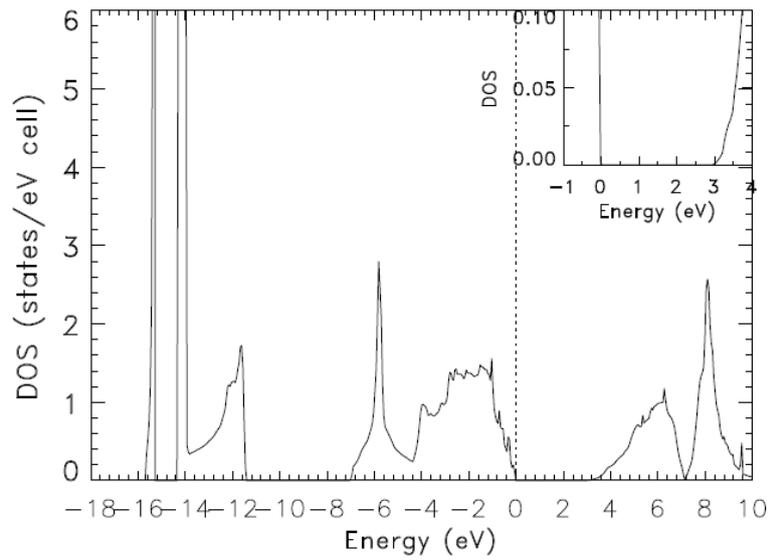

**Figure 2b.** Calculated, total density of state (DOS) for zb-GaN, as derived from the bands from calculation III, as shown in Fig.1b

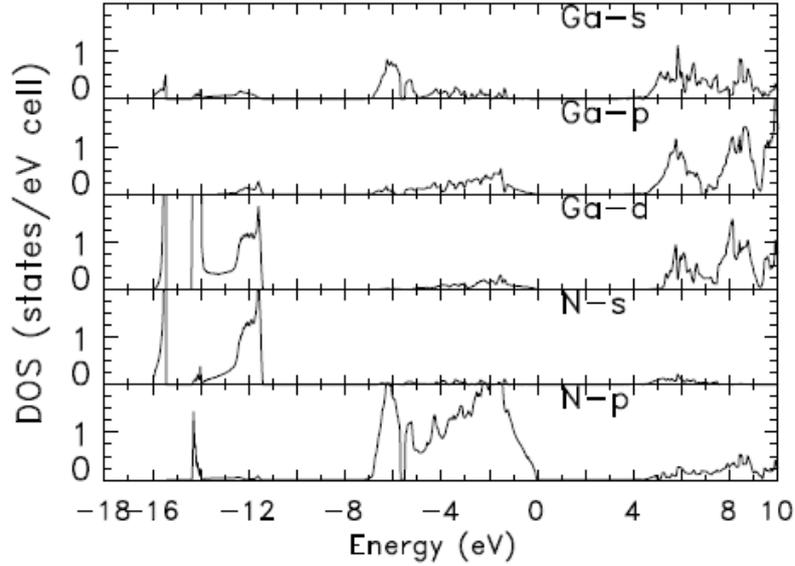

**Figure 3a**. Calculated, partial density of state (pDOS) for w-GaN, as derived from the bands from calculations III, in Fig. 1a

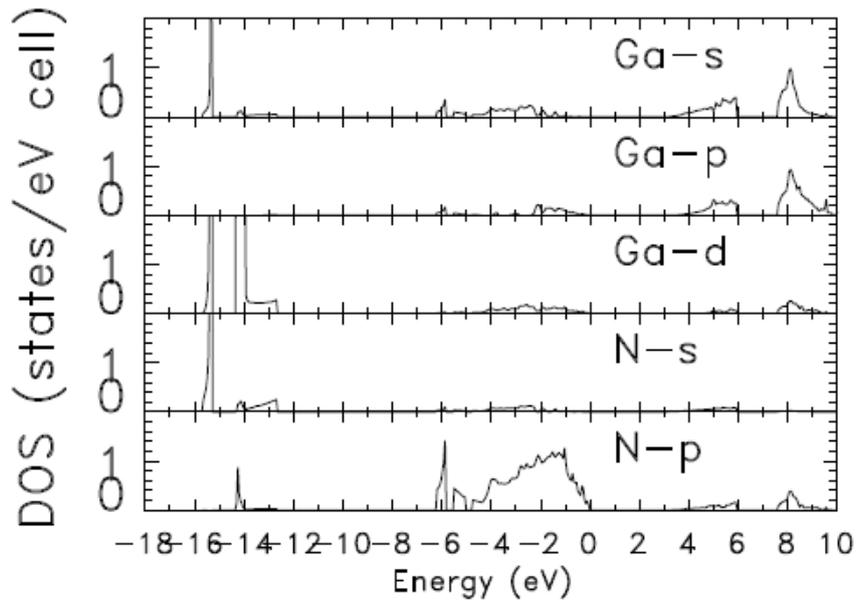

**Figure 3b**. Calculated, partial density of state (pDOS) for zb-GaN, as derived from the bands from calculations III, in Fig. 1b

The significance of the above bands and related densities of states for the correct description of GaN is underscored by the following points. In contrast to most previous ab initio DFT calculations in Tables 1a and 1b, our results for both w-GaN and zb-GaN practically agree with experiment for the band gaps. Given the importance of the band gap in most applications of semiconductors, accurately calculated band gaps have the potential of saving time and money presently spent in the trials and errors needed for semiconductor based device design and fabrication. Compared to most of the previous DFT calculations, densities of states and dielectric functions calculated with the bands and wave functions from DFT BZW-EF calculations generally agree with experiment, as shown for BaTiO$_3$ [24] and wurtzite InN [37]. Specifically, once the calculated band gap is significantly off the experimental values, peaks in plots of calculated dielectric functions will be off the corresponding experimental ones. The same is true for the peaks in the total and partial densities of states for the conduction bands.

Our results shed some light on some difficulties associated with calculations with LDA and GGA potentials or various modified DFT potentials partly developed with the aim of resolving the band gap problem. The bands for zb-GaN, in Fig. 1b, show that the lowest valence bands, d bands, are clearly not core states, due to the dependence of the energy on the k vector, however small it may be. Qteish et al. [38] performed LDA calculations for zb-GaN, with the d electrons in the core and in the valence, to obtain band gaps of 2.20 eV and 1.65 eV, respectively. While the former value, obtained with the d electrons in the core, is much colder to the experimental gap of 3.2 eV, the content of Fig. 1b suggests that it is not the acceptable one. The exact exchange (EXX) and LDA-GW calculations by Qteish et al., for zb-GaN, led to gap of 2.67 and 2.88 eV, respectively, with the d electrons in the valence. These values are much closer to the experimental one. The 2.88 eV LDA-GW result is about the same as our LDA BZW-EF one of 2.896 (i.e. 2.90) eV. This result, like many others from our group, indicates that ab initio DFT calculations can produce the correct description of materials without any adjustments, provided that said calculations adhere to conditions that are inherent to derivation, and hence validity, of DFT.

The extensive X-ray spectroscopy measurements and first principles calculations by Magnuson et al. [29] allow informative comparisons of their findings with our results. Their DFT calculations with the Wu and Cohen GGA potential and the WIEN2K code is reported to have led to a band gap of 1.83 eV for w-GaN. The serious underestimation of the experimental value (3.26-3.41 eV) is understandable, given that their calculations presumably utilized a single basis set approach. The same calculations have placed the d bands at -13.5 eV below the valence band maximum, as opposed to the experimental, binding energy of -17.5 0.1 eV. These authors corrected the binding energy problem with a WG-GGA+U calculation, with U equals to 10 eV. Our calculated band gap of 3.29 eV for w-GaN is 1.46 eV larger than the 1.83 eV found by Magnuson et al. As for the binding energy, we found -15.968 ( that is much closer to the experimental value given above. We should note that Ley et al. [39] reported binding energy measurements for GaP where the X-ray and ultraviolet data differed by 1.1 eV in absolute value, with the X-ray result having the largest absolute value. Similar differences were reported for other materials, i.e., 0.8 eV for GaAs. While we do not claim that one result is better than the other, our calculated binding energy could be much closer to one from UPS measurements.

While our results differ significantly from some theoretical ones of Ley et al. [39] as discussed above, our partial densities of states for w-GaN agree with several of their experimental findings relative to the hybridization of s, p, and d states as illustrated in the following. As per Fig. 3a above, we find the N-p and Ga-p hybridization between -1 eV and -2 eV, exactly as Ley et al. reported it. Similarly, N 2p and Ga 4s hybridize between -5.5 and -6.5 eV in Fig. 3a as reported by Ley et al. Ley et al. While Ley et al. found the hybridization of N 2p, N2s, and Ga 4s between -13 and -15 eV, we found Ga 3d to dominate in this energy range, with a significant contribution from N 2p, a small one from N 2s and a feint one from Ga 4s.

**Table 4a.** Calculated, electronic energies of wurtzite gallium nitrite (w-GaN) at high symmetry points in the Brillouin zone, as obtained with the optimal basis set (Calculation III). The energies are in electron volts (eV) and the value of 0.0 eV indicates the position of the Fermi level. The calculated, direct band gap is 3.289 ( 3.29) eV, at the Γ point.

| A-point | L-point | M-point | Γ-point | H-point | K-point |
|---:|---:|---:|---:|---:|---:|
| 9.099 | 8.221 | 9.261 | 10.192 | 7.656 | 8.718 |
| 9.099 | 8.221 | 6.920 | 10.065 | 7.656 | 8.633 |
| 5.269 | 4.767 | 5.972 | 5.088 | 6.803 | 8.633 |
| 5.269 | 4.767 | 5.334 | **3.289** | 6.803 | 5.391 |
| -0.574 | -1.952 | -1.051 | **0** | -1.538 | -2.804 |
| -0.574 | -1.952 | -2.321 | -0.030 | -1.538 | -3.153 |
| -0.574 | -2.060 | -2.907 | -0.030 | -4.354 | -3.153 |
| -0.574 | -2.060 | -4.372 | -1.115 | -4.354 | -5.340 |
| -3.825 | -6.731 | -5.306 | -1.115 | -6.254 | -5.340 |
| -3.825 | -6.731 | -6.546 | -7.243 | -6.254 | -5.499 |
| -13.829 | -12.054 | -11.674 | -12.203 | -11.702 | -11.724 |
| -13.829 | -12.054 | -12.354 | -13.993 | -11.702 | -11.724 |
| -14.035 | -13.999 | -13.968 | -13.993 | -14.051 | -14.012 |
| -14.035 | -13.999 | -14.009 | -14.094 | -14.051 | -14.063 |
| -14.035 | -14.067 | -14.090 | -14.094 | -14.073 | -14.110 |
| -14.035 | -14.067 | -14.099 | -14.294 | -14.073 | -14.110 |
| -14.334 | -14.182 | -14.128 | -14.294 | -14.212 | -14.128 |
| -14.334 | -14.182 | -14.178 | -14.355 | -14.212 | -14.151 |
| -14.334 | -14.301 | -14.279 | -14.355 | -14.236 | -14.151 |
| -14.334 | -14.301 | -14.300 | -14.374 | -14.236 | -14.348 |
| -15.633 | -15.446 | -15.459 | -15.519 | -15.426 | -15.473 |
| -15.633 | -15.446 | -15.525 | -15.968 | -15.426 | -15.473 |

**Table 4b.** Calculated, electronic energies of zb-GaN at high symmetry points in the Brillouin zone, as obtained with the optimal basis set of Calculation III. The used, experimental lattice constant is 4.50 Å. The Fermi energy is set to zero. The calculated, direct band gap, at Γ, is 2.896 eV (2.90 eV).

| L-POINT | Γ-POINT | X-POINT | K-POINT |
|---|---|---|---|
| 10.466 | 10.618 | 14.104 | 13.000 |
| 10.466 | 10.618 | 12.064 | 11.536 |
| 9.068 | 10.618 | 7.065 | 8.301 |
| 5.545 | **2.896** | 3.400 | 5.092 |
| -0.996 | **0** | -2.768 | -2.224 |
| -0.996 | **0** | -2.768 | -4.327 |
| -7.052 | **0** | -6.074 | -5.795 |
| -12.238 | -14.043 | -11.681 | -11.554 |
| -14.003 | -14.043 | -13.973 | -14.008 |
| -14.003 | -14.353 | -14.072 | -14.105 |
| -14.268 | -14.353 | -14.072 | -14.115 |
| -14.268 | -14.353 | -14.249 | -14.229 |
| -15.374 | -15.700 | -15.331 | -15.325 |

The tables above are intended to allow extensive comparisons of our results with future, experimental ones. These experiments include angle resolved photoemission, X-ray, and ultraviolet spectroscopy and various optical measurements. Readily available from these tables are the widths of the various groups of valence bands and the total widths of the valence bands. From Table 4a, the widths of the lowest laying, middle, and upper most groups of valence bands for w-GaN are 0.542, approximately 2.700, and 7.243 eV, respectively. The total valence band width is 15.968 (     eV. Similarly, from Table 4b., the widths of the lowest laying, middle, and upper most groups of valence bands for zinc blende GaN, as well as the total band valence band width, are 0.375, 2.799, 7.051, and 15.700 eV, respectively. These tables readily allow the calculation of several specific, transition energies, including optical ones.

## b. Discussions

Even though the bands from Calculations III and IV are undistinguishable for the occupied energies and, up to 10 eV, for the unoccupied ones, that is not the case for much higher, unoccupied energies. For the latter, some unoccupied energies from Calculations IV are lower than corresponding ones from Calculations III, for both w-GaN and zb-GaN. The differences between these unoccupied, higher energies are far from being the same, indicating that the higher energy bands from Calculations IV are not rigidly shifted downward as compared to those from Calculations III. The above superposition of energies up to 10 eV explains the agreement between our calculated, optical band gaps and experimental ones, taking into account possible Burstein Moss, sample quality, film thickness, and temperature effects.

The stark difference between results of ab initio DFT calculations in Table 1a and 1b and the ones we obtained naturally poses the question as to what differences between these calculations are ours could be the explanation. The answer to this question is partly provided in the description of our BZW-EF method, inasmuch as the implementation of that method is the major difference between our work and the others. Indeed, most of the previous ab initio and self consistent calculations utilized a single basis set, judiciously selected, to perform iterations that led to stationary solutions. Even if some calculations utilized more than one basis set, we are not aware of any information suggesting that successively embedded basis sets were employed, as done by the BZW-EF method, to verifiably attain the absolute minima of the occupied energies. Hohenberg and Kohn, [40] through the variational principle for DFT, also known as the second Hohenberg-Kohn theorem, asserted the following: *For the results of ab initio self consistent calculations using a DFT potential to provide th ground state description of a system, it is necessary either (a) to have, à priori, the "correct" ground state charge density as input or (b) to obtain the minimum of the energy content of the Hamiltonian as a functional of the charge density*. In the former case, (a), the energy obtained with the "correct" charge density is the ground state energy; in the latter case, (b), the charge density that leads to the minimum of the minimum of the energy functional is the ground state density.

In light of the preceding, the reason for the agreement between our calculated band gaps and other properties and corresponding, experimental ones resides in our strict adherence to conditions that are inherent to the derivation (and hence validity) of DFT.

## IV. Conclusion

We have performed ab-initio, self-consistent calculations of electronic energy bands, total (DOS), and partial (pDOS) densities of states (DOS) of w-GaN and zb-GaN. The implementation of the Bagayoko, Zhao, and Williams method, as enhanced by Ekuma and Franklin (BZW-EF), is the key difference between our approach and that in most of the previous ab initio calculations.. Our calculated band gaps of 3.29 eV and 2.9 eV for w-GaN and zb-GaN, respectively, are in basic agreement with corresponding, experimental ones. These gaps, obtained at room temperature lattice constants, are significant improvements over most of the previous ab initio DFT results. The only distinctive feature of our calculations as compared to most of the previous ones stems from our strict implementation of the LCAO formalism using the Bagayoko, Zhao, and Williams method as enhanced by Ekuma and Franklin (BZW-EF). Our findings and the related notes at the end of the discussion above point to a need to revise some limitations previously ascribed to DFT and particularly to DFT potential. The DFT variational principle needs to be heeded by calculations before their results can be fully credited to DFT.

## Acknowledgements

Research work funded by the Malian Ministry of Higher Education and Scientific Research, through the training of trainers program (TTP), the Ministry of Expatriate Malians, through the "Transfer of Knowledge Through Expatriate National" (TOKTEN) project, the US National Science Foundation [NSF, Award Nos. EPS-1003897, NSF (2010-15)-RII-SUBR, and HRD-1002541], the US Department of Energy, National Nuclear Security Administration (NNSA, Award No. DE-NA0001861), and LONI-SUBR.